\documentclass{article}
\usepackage{spconf,amsmath,graphicx}

\usepackage{makecell}
\usepackage{multirow}
\usepackage{comment}

\title{Attention or Convolution: Transformer Encoders in Audio Language Models for Inference Efficiency}

\name{Sungho Jeon$^{1*}$\thanks{*This work has done while Sungho Jeon was interning at Meta}, Ching-Feng Yeh$^{2}$, Hakan Inan$^{2}$, Wei-Ning Hsu$^{2}$, Rashi Rungta$^{2}$, Yashar Mehdad$^{2}$, \obeylines \textit{Daniel Bikel$^{2}$} }

\address{
  $^{1}$Heidelberg Institute of Theoretical Studies \\
  $^{2}$Meta
}

\begin{document}

\maketitle

\begin{abstract}

In this paper, we show that a simple audio language model can achieve comparable inference efficiency to more complicated pre-trained models with speech transformer encoders. These speech transformers rely on mixing convolutional modules with self-attention modules. They achieve state-of-the-art performance on ASR with top efficiency. We first show that employing these speech transformers as an encoder significantly improves the efficiency of audio language models as well. However, our study shows that we can achieve comparable efficiency with advanced self-attention solely. We demonstrate that this simpler approach is particularly beneficial with a low-bit weight quantization technique of a neural network to improve efficiency. We hypothesize that it prevents propagating the errors between different quantized modules compared to recent speech transformers mixing quantized convolution and the quantized self-attention modules. Our study suggests that we could pay attention to the architecture of audio language language to improve their inference efficiency.

\end{abstract}


\section{Introduction}

Self-supervised audio language models exploit unlabeled data to learn audio representations, agnostic to a specific task. These representations are used for the target task by fine-tuning instead of supervised learning with massive amounts of labeled data. This paradigm of pre-training and then fine-tuning alleviates the dependency on abundant labeled data, and it lets us deploy our AI systems for diverse problems more easily. Following this paradigm, pre-training frameworks exploit rather complicated architectures \cite{schneider2019wav2vec}. These frameworks bring larger parameter counts with the concomitant higher computational cost of inference.

However, there is a large gap between this high computational inference cost and the requirements for deploying these models for on-device problems such as automatic speech recognition (ASR) for wearable devices. To narrow this gap, recent work mostly investigates different configurations for the components in the architecture, such as efficient configurations for feature extractors \cite{wu2022performance} or subsampling of input sequences \cite{vyas2022demand}. Another line of research investigates the efficient transformer itself designed for the target audio task solely. Gulati et al.~\cite{gulati2020conformer} introduce a transformer mixing a convolutional and a self-attention module. Their supervised model performs better with a smaller model size on a standard ASR dataset than pre-trained audio language models. Inspired by this transformer, a more efficient model was introduced \cite{kimsqueezeformer}. These modern speech transformers benefit from using convolutional modules in conjunction with self-attention modules. 

Interestingly, this approach is different from recent efficient model architectures studied in other areas of AI. In the area of Natural Language Processing (NLP), an efficient transformer has been studied mainly by introducing more efficient components of a vanilla transformer without mixing convolutional modules \cite{child2019generating}. In the area of Computer Vision, Dosovitskiy et al.~\cite{dosovitskiyimage} focus on the training setup for the tasks of image recognition. They treat image patches in the same way as textual items are treated in NLP. This work shows that a simple transformer can achieve comparable performance with the state-of-the-art models mixing convolutional modules with self-attention modules. More recently, a similar artifact is shown on the speech tasks as well \cite{gong2022ssast}. However, this study focuses on the perspective of performance rather than the perspective of efficiency.

\begin{figure*}
    \centering
    \includegraphics[width=0.7\textwidth]{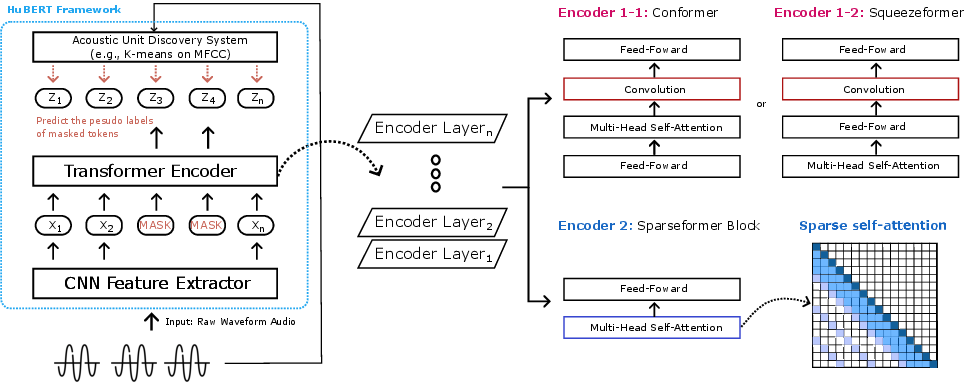}
    \caption{HuBERT framework and two types of encoder candidates: Conformer or Squeezeformer vs. Sparseformer.}
    \label{fig:archi_model_combine}
\end{figure*}

In this work, we investigate the inference efficiency trade-offs of the transformer encoder, employed in the self-supervised audio language models. We first show that we can improve the inference efficiency of audio language models by employing modern speech transformers ---Conformer and Squeezeformer--- as their encoder. It improves their inference efficiency significantly, with comparable performance at lower cost. However, our study shows that we can achieve comparable efficiency with only efficient self-attention, without mixing convolution modules. 

Our evaluation shows that this approach is particularly beneficial when we apply a quantization technique to improve efficiency. This approach with quantization reduces 93.4\% of the storage size, more than 90\% of the computational cost but degrades the performance on the 10 downstream tasks such as increasing word error rate from 6.89\% to 19.33\% in ASR, compared to the original pre-trained model without quantization. We hypothesize that a simple transformer prevents propagating errors between different quantized modules compared to modern speech transformers that mix modules of differing types.

\section{Related work}

One line of related research investigates the efficiency of audio language models from the perspective of pre-processing audio data. Wu et al.~\cite{wu2022performance} investigate the architecture variations of Wav2Vec 2.0 framework, a self-supervised pre-training framework \cite{baevski2020wav2vec}, to examine the inference efficiency trade-offs. They propose several techniques to improve the efficiency of Wav2Vec 2.0 framework. For example, they introduce more efficient configurations for the feature extractor and downsampling input sequences linearly before their Transformer encoder. Following this work, Vyas et al.~\cite{vyas2022demand} propose a stochastic approach to sub-sampling input sequences. 
Nevertheless, there is little attention on the influence of their Transformer encoder in terms of efficiency. 

Since earlier studies of audio language models are based on a vanilla Transformer, recent studies investigate the influence of a more advanced Transformer encoder. Zhang et al.~\cite{zhang2020push} employ Conformer into Wav2Vec 2.0, and this approach improves the performance on audio downstream tasks. 
Instead of deploying Conformer, Chen et al.~\cite{chen2022wavlm} propose a masked speech denoising and a pre-training framework which employs a Transformer encoder with relative positional encoding. Their pre-trained model outperforms a HuBERT model. However, previous work mostly focuses on the perspective of performance but there has been little interest on the efficiency on this line of research.

Another line of studying the efficiency of neural networks is quantizing the components of neural networks \cite{hubara2016binarized}. Earlier studies replace all full-precision weights of a neural network with lower-precision weights. This approach drastically reduces the memory size and inference time. However, quantizing the weights of a whole network can cause the propagation of errors between modules, and the accumulated errors degrade the performance significantly in the end. To alleviate this, diverse techniques have been introduced including a partial quantization of weights. 
More recently, a binarized Transformer is proposed, which employs a learnable scaling method to the lower bits \cite{liubit2022}. Yet et al.~\cite{yeh2022efficient} show that pre-trained audio models can benefit from this work as well. 
Following this work, we investigate the inference efficiency trade-offs with a quantization of neural network weights.

\begin{table*}[t]
	\centering  \scriptsize 
	
	\begin{tabular}{c || c || c c  c  c c c || c c c}
            
          \Xhline{2\arrayrulewidth}
            Encoder in HuBERT & Encoder Params & Prec & Storage & FLOP & BOP\_NonQ & BOP\_BQ & NVDA\_Est\_Time & ASR$\downarrow$ & SD$\downarrow$ \\
             (+FastConv) & L / D / H & & (MB) & (Gs) & (Gs) & (Gs) & (e-04, second) & (WER, \%) & (DER, \%) \\ \Xhline{2\arrayrulewidth}
            
             Vanilla Trans  & 12 / 786 / 12 & FP32 & 184.42 & 110.54 & 1228.64 & - & 38.46 & 7.06 & 6.32 \\ \hline
            \textbf{Conformer-S} & 16 / 144 / 4 & FP32 & 131.87 & 22.10 & 329.39 & - & 10.31 &  8.56 & 6.81 \\
           Squeeze-XS & 16 / 144 / 4 & FP32 & 132.04 & 18.31 & 272.88 & - & 8.54 & 8.96 & 9.18 \\
           \hline
           
          \textbf{Sparseformer-DN-S} & 16 / 256 / 4 & FP32 & 60.81 & 26.05 & 388.18 & - & 12.15 & 8.44 & 7.66 \\
          Sparseformer-SW-S & 8 / 512 / 4 & FP32 & 117.18 & 40.09 & 597.43 & - & 18.70 & 7.88 & 6.56 \\ 
           
           \hline
           \hline
                      
           BQ-Vanilla Trans  & 12 / 786 / 12  & FP32-W1A1 & 25.23 & 11.82 & 172.83 & 63.62 & 5.53 & 16.83 & 7.62 \\ \hline
           \textbf{BQ-Conformer-S }& 16 / 144 / 4 & FP32-W1A1 & 12.88 & 7.23 & 103.44 & 15.94 & 3.27 & 20.52 & 11.11 \\
           BQ-Squeeze-XS & 16 / 144 / 4  & FP32-W1A1 & 13.05 & 7.20 & 104.05 & 11.86 & 3.28 & 24.10 & 13.42 \\
           \hline
           \textbf{BQ-Sparseformer-DN-S} & 16 / 256 / 4 & FP32-W1A1 & 12.10 & 7.35 & 107.91 & 10.70 & 3.40 & 19.33 & 9.64 \\ 
           BQ-Sparseformer-SW-S & 8 / 512 / 4  & FP32-W1A1 & 19.71 & 8.85 & 130.38 & 19.49 & 4.12 & 18.15 & 8.39 \\ 
             
           \Xhline{2\arrayrulewidth}
            
        \end{tabular}

    \caption{Profiling Results (L: Layer Num, D: Dim, H: Head Num; BQ: BiT Quantization W1A1; NVDA\_Est\_Time: Estimated time for their BOP based on the catalog of NVidia A100)}
	\label{tab:profile}
\end{table*}

  




\section{Model Architecture}


\subsection{Self-Supervised Audio Pre-Training: HuBERT}
Our study is based on HuBERT \cite{hsu2021hubert} for a framework of self-supervised audio pre-training (Figure \ref{fig:archi_model_combine}). This framework consists of three components: a feature extractor, a transformer encoder, and an acoustic unit discovery module. Following the Wav2Vec 2.0 architecture, a convolutional waveform component is employed for a feature extractor which takes raw waveform inputs. It projects the input to vector representations. The transformer encoder consists of multiple blocks, and it processes the input representations. The acoustic unit discovery module produces the pseudo-labels of input audio frames by clustering features, such as clustering MFCC features via $k$-means. Inspired by the BERT pre-training, non-masked audio representations are learned to describe the masked tokens well by predicting their pseudo-labels.

\subsection{Encoder Candidate 1: Conformer / Squeezeformer}

Conformer, the convolution-augmented transformer, consists of stacked layers of convolutional modules in conjunction with self-attention modules. It achieves comparable performance on ASR with a smaller model size compared to a vanilla transformer. Conformer is originally designed for ASR, but it has been used widely for an efficient audio transformer in other speech tasks as well.

Kim et al.~\cite{kimsqueezeformer} redesign the Conformer architecture based on their empirical study, with a new architecture they call Squeezeformer. They investigate two aspects, at the micro level and at the macro level. For the macro level, they introduce subsampling of input audio sequences. 
For the micro-level, they introduce several modifications including re-ordering the modules in the transformer, changing activation functions, and reducing the number of layer normalization modules.

\subsection{Encoder Candidate 2: Sparseformer}

Local window attention has been studied to deal with long input sequences. The full-attention matrix is sparsified by attention patterns, which scales linearly for the input sequence length. Following this, Sparseformer achieves similar performance with a vanilla transformer with significantly fewer operations \cite{child2019generating}. The key idea of Sparseformer is to subdivide a full-attention computation into several sub-computations first, which applies a fixed-attention pattern as hyper parameters. Then these sub-computed outputs are used to approximate the full-attention.

\subsection{Neural Quantization: Robustly Binarized Transformer}

Liu et al.~\cite{liubit2022} propose the robustly binarized transformer (BiT), which is a fully binarized transformer. They introduce a two-set binarization scheme and an elastic binarization function which learns the mapping range of quantization in the training. We employ this quantization technique to investigate the influence of different transformer encoders with quantization. While Liu et al.~\cite{liubit2022} focus on quantizing a transformer and linear/activation layers, we implement their quantization techniques for the convolutional layers as well to quantize Conformer and Squeezeformer. Yeh et al.~\cite{yeh2022efficient} investigate the influence of different target bits for quantizing the HuBERT-base model with a vanilla transformer. We only investigate the extreme bit of quantization, both 1 bit for weights and activation (W1A1).

\begin{table*}[t]
	\centering \scriptsize 
            \begin{tabular}{c | c || c | c | c | c | c | c | c | c | c | c}
            
          \Xhline{2\arrayrulewidth}
            Encoder in HuBERT & \multirow{2}{*}{Prec} & \multicolumn{10}{c}{SUPERB Tasks} \\
             (+FastConv) & & ASR$\downarrow$ & KS$\uparrow$ & SF$\uparrow$ & SID$\uparrow$ & PR$\downarrow$ & QbE$\uparrow$ & IC$\uparrow$ & ASV$\downarrow$ & SD$\downarrow$ & ER$\uparrow$ \\ \Xhline{2\arrayrulewidth}
            
             Vanilla Trans (Reported)  & FP16 & 7.06 & 96.62 & 0.89 & 53.67 & 6.05 & 6.91 & 97.28 & 5.30 & 6.32 & 65.00 \\
             Vanilla Trans (Our Setup) & FP32 & 6.89 & 96.56 & 0.89 & 53.52 & 5.82 & 6.70 & 97.86 & 5.96 & 6.63 & 63.83 \\ \hline
            \textbf{Conformer-S} & FP32 & 8.56 & 95.94 & 0.87 & 52.11 & 9.42 & 6.20 & 92.88 & 5.80 & 6.81 & 62.12 \\
           Squeeze-XS & FP32 & 8.96 & 96.13 & 0.80 & 35.35 & 10.73 & 6.14 & 83.31 & 7.86 & 9.18 & 58.75 \\
           \hline
           
          \textbf{Sparseformer-DN-S} & FP32 & 8.44 & 95.89 & 0.88 & 60.57 & 7.99 & 6.27 & 96.02 & 6.24 & 7.66 & 62.26 \\
          Sparseformer-SW-S & FP32 & 7.88 & 93.25 & 0.88 & 61.86 & 9.72 & 4.92 & 93.15 & 6.98 & 6.56 & 65.90\\
           
           \hline \hline


           
           BQ-Vanilla Trans (Reported) & FP16-W1A1 & 15.96 & 93.83 & 0.78 & 49.62 & 22.96 & 5.63 & 93.01 & 6.83 & 7.62 & 61.68 \\
           BQ-Vanilla Trans (Our Setup) & FP32-W1A1 & 16.83 & 94.77 & 0.79 & 40.15 & 20.63 & 5.57 & 89.77 & 9.13 & 7.80 & 60.74 \\
            \hline

           \textbf{BQ-Conformer-S} & FP32-W1A1 & 20.53 & 92.44 & 0.76 & 24.98 & 37.13 & 5.18 & 69.73 & 9.86 & 11.11 & 56.41 \\
           BQ-Squeeze-XS & FP32-W1A1 & 24.10 & 92.79 & 0.69 & 18.56 & 28.82 & 5.08 & 62.01 & 11.97 & 13.42 & 57.57 \\ \hline
           \textbf{BQ-Sparseformer-DN-S} & FP32-W1A1 & 19.33 & 92.24 & 0.79 & 29.21  & 33.17 & 4.66 & 71.34 & 10.70 & 9.64 & 58.97 \\
           BQ-Sparseformer-SW-S & FP32-W1A1 & 18.15 & 94.03 & 0.79 & 38.92 & 24.37 & 6.09 & 84.37 & 9.19 & 8.39 & 61.39\\
           \Xhline{2\arrayrulewidth}
            
        \end{tabular}

\caption{Evaluation on SUPERB Tasks (BQ: BiT Neural Quantization, W1A1: 1 bit for Weights and Activation, Reported: Reported in \cite{yeh2022efficient}, Our Setup: Lower batch size (0.5M tokens $<$ 1.2M tokens))}
	
		\label{tab:superb_all}
\end{table*}

\section{Experiments}
\subsection{Experimental Setup}
\noindent\textbf{Pre-training setup.} We follow the pre-training setup of HuBERT. This pre-training is based on the Librispeech dataset, consisting of 960 hours. 
We use 32 GPUs of NVidia A100 with a batch size of at most 36.5 seconds of audio per GPU. All models are trained for 250k steps in the first phase, then they are trained for 600k steps in the second phase. It takes 8.5 hours for 100k steps on our setup.

\bigskip

\noindent\textbf{Evaluation Setup.}
We evaluate models in terms of computational cost and performance on downstream tasks. We first profile the computational cost of models using the DeepSpeed library \cite{rasley2020deepspeed}. We examine a required storage (Storage), a number of floating point operations (FLOP), a number of bit operations (BOP) \cite{van2020bayesian}, and an estimated time for their BOP based on a catalog of NVidia A100 GPU. 

We evaluate models on the 10 downstream tasks of SUPERB \cite{yang21c_interspeech}: automatic speech recognition (ASR), keyword spotting (KS), slot filling (SF), speaker identification (SID), phoneme recognition (PR), query by example (QbE), intent classification (IC), automatic speaker verification (ASV), speaker diarization (SD) and emotion recognition (ER). 

\bigskip

\noindent\textbf{Model configurations.}
We employ the setup of baselines' smallest model, Conforemr-S and Squeezeformer-XS, respectively \cite{gulati2020conformer,kimsqueezeformer} (Table \ref{tab:profile}). Following their shape of deep-narrow architecture, we design Sparseformer-DN-S which requires smaller computational cost than others in the quantized models

\subsection{Efficiency: Model Profiling and Downstream Tasks}

\noindent\textbf{Inference efficiency trade-offs.} We first compare the computational cost of different encoders without quantization (Table \ref{tab:profile}). A model employing Conformer-S shows lower cost than the baseline, 64\% for the required storage and 74\% in FLOP reductions compared to the HuBERT with vanilla Transformer. Since Squeezeformer has the same fundamental architecture with Conformer, it shows similar profiling results. Sparseformer-DN-S also shows comparable reductions for their computational cost. 

Next, we evaluate these models on the 10 speech downstream tasks of SUPERB (Table \ref{tab:superb_all}). We observe the efficiency trade-offs for employing more efficient transformer on downstream tasks. For example, it increases word error rate from 6.89 to 8.44 in ASR.

\bigskip

\noindent\textbf{Efficiency with quantization.} When we apply 1-bit BiT quantization, our results show that these two types of encoders have different influences. The quantized model employing Spareformer (BQ-Sparseformer-DN-S) shows better performance than the quantized model employing Conformer-S (BQ-Conformer-S) overall. It shows a lower word error rate on ASR (19.33 $<$ 20.52) and a lower diarization error rate on SD (9.64 $<$ 11.11). Despite of the fact that BQ-Sparseformer-DN-S takes the smallest computational cost compared to the cost of BQ-Conformer-S: the 7.7\% smaller required storage and the 32.9\% smaller BOP\_BQ. We hypothesize that the quantized modules of different types in these speech transformers propagate errors. Then, the accumulated errors degrade performance more than a simple transformer encoder, consisting of self-attention modules only.

Compared to the baseline without quantization, employing Sparseformer with BiT quantization reduces 93.4\% required storage (184.42 $\rightarrow$ 12.10), 93.4\% of FLOP (110.54 $\rightarrow$ 7.35 ), and 90.3\% of BOP (1228.64 $\rightarrow$ 118.61). We estimate 91.1\% runtime reduction in the theoretical maximum performance of an NVidia A100 GPU. In return, it increases the word error rate from 6.89\% to 19.33\%, and other tasks as well. Compared to the baseline with BiT quantization, it saves 52.1\% of required storage (25.23 $\rightarrow$ 12.10), 37.8\% of FLOP (11.82 $\rightarrow$ 7.35), 50\% of BOP (236.45 $\rightarrow$ 118.61). As efficiency trade-offs, it increases word error rate from 16.83\% to 19.33\%, and overall.

\subsection{Architecture Shape: Deep-Narrow vs. Shallow-Wide}
\label{subsec: archi_shape}
Ashihara et al.~\cite{ashihara2022deep} show that two different shapes of architectures have different advantages as speech tasks when they investigate this issue with knowledge distillation. Inspired by this, we design a shallow-wide shape of Sparseformer (Sparseformer-SW-S). It has half of the number of layers but twice larger dimensions compared to the setup of Sparseformer-DN-S. This model shows better performance, but this shape of architecture brings higher computational costs. Our profiler shows that the wide shape of the architecture causes a larger storage size due to the absolute positional encoding, employed in the framework. Each layer also requires more matrix multiplication operations due to the self-attention mechanism. Hence, it has disadvantages in computational cost to design more efficient models.

\section{Conclusions}
We investigate the efficiency trade-offs of employing different transformer encoders into the self-supervised framework of audio pre-training. Our experiments show that there are decent efficiency trade-offs when we employ them. When we apply a quantization technique, however, our results suggest that a simple transformer encoder employing only efficient self-attention modules is more beneficial than the recent speech transformers blending modules of differing types.




\bibliographystyle{IEEEbib}
\bibliography{mybib}


\end{document}